# Title

Efficient monitoring of blood-stage infection in a malaria rodent model by the rotating-crystal magneto-optical method

# Authors

Agnes Orban[1], Maria Rebelo[2], Inês S. Albuquerque[2], Adam Butykai[1], Istvan Kezsmarki[1], Thomas Hänscheid[2]

# Affiliations

1 Department of Physics, Budapest University of Technology and Economics and MTA-BME Lendület Magneto-optical Spectroscopy Research Group, 1111 Budapest, Hungary

2 Instituto de Medicina Molecular, Faculdade de Medicina Universidade de Lisboa, 1649-028 Lisbon, Portugal

# Corresponding Author

Agnes Orban: orbanag@gmail.com

# Authors' contributions

AO, MR and ISA performed the experiments. AO, MR, ISA and IK analyzed the data. AO and MR wrote the manuscript. AO and AB developed the magneto-optical setup. TH an IK designed the experiments and supervised the project. All authors read and approved the final form of the manuscript.

# Competing Interests

The authors declare that they have no competing interests.

# Acknowledgements

## Abstract

Intense research efforts have been focused on the improvement of the efficiency and sensitivity of malaria diagnostics, especially in resource-limited settings for the detection of asymptomatic infections. A recently developed magneto-optical (MO) method allows for the high-sensitivity detection of malaria pigment (hemozoin) crystals in blood via their magnetically induced rotational motion. First evaluations of the method using synthetic β-hematin crystals and *in vitro P. falciparum* cultures imply its potential for in-field diagnosis. To further investigate this potential, here we study the performance of the method in monitoring the *in vivo* onset and progression of the blood stage infection in a rodent malaria model. We found that the MO method can detect the first generation of intraerythrocytic parasites at the ring stage 61-66 hours after sporozoite injection, demonstrating better sensitivity than light microscopy or flow cytometry. MO measurements performed after the treatment of *P. berghei* infections show that the clearance period of hemozoin in mice is approx. 5 days, which indicates the feasibility of the detection of later reinfections as well. Being label and reagent-free, cost-effective and rapid, together with the demonstrated sensitivity, we believe that the MO method is ready for in-depth clinical evaluation in endemic settings.

## Introduction

As malaria remains one of the most serious health and economic burdens of the globe, the study of human malaria is one of the most important agendas of today's scientific research. The study of the disease involves a myriad of methods ranging from epidemiological analysis and clinical studies to laboratory model systems such as *in vitro* parasite cultures in human red blood cells (RBCs) and *in vivo* rodent models. Rodent models of malaria have been widely and successfully used to study the biology and pathology of the malaria infection and host immune response [reviewed in Zuzarte-Luis 2014a], as well as for *in vivo* evaluation of novel vaccine and drug candidates [Mota 2001, Khan 1991], despite their limitations to replicate certain aspects of the human disease [White 2012, Craig 2008].

In such studies the most commonly used parameter to monitor the progression of the infection is the appearance of parasites in the circulation and the assessment of parasitemia, which is often challenging to rapidly, yet accurately determine.

The reference method of light microscopy of Giemsa-stained thin blood smears is a demanding, labor-intensive procedure, often lacking the desired sensitivity, especially at low parasite densities usually characterizing the early time points of intraerythrocytic development [Lelliot 2014]. On the other hand, assays using molecular methods (PCR-based techniques) that far surpass the performance of light microscopy, are still not used for the continuous monitoring of the infection in mouse experiments as they are time consuming and expensive techniques [Valkiunas 2008, Taylor 2012].

Automated approaches, such as flow cytometry are preferred and extensively used, however, to achieve sufficient sensitivity for the detection of early blood-stage infections they often require special dyes combined with complex protocols [Lelliot 2014, Jimenez-Diaz 2009, Malleret 2011] or transgenic green fluorescent protein (GFP) expressing parasites [Franke-Fayard 2004, Franke-Fayard 2008]. These transgenic luciferase expressing lines have also been demonstrated to be appropriate subjects of a new and simple bioluminescence assay for the accurate evaluation of the pre-patent period, i.e., the time between sporozoite inoculation and the appearance of parasites in the peripheral circulation, and the early blood stages [Zuzarte-Luis 2014b]. However, methods using the chemiluminescent properties have certain limitations, as so far they have been restricted to two species of rodent malaria, *P. yoelii* [Ono 2007] and *P. berghei* strain ANKA [Franke-Fayard 2004], and the luciferin substrast used for these assays is costly.

The need for a universally applicable and automated method for blood stage parasite quantification has motivated several investigations of the unique physical properties of malaria pigment (hemozoin-hz). As a result, its utilization as an alternative target of magnetic and/or optical detection of the infection has been proposed by several groups [Rebelo 2012, Rebelo 2015, Karl 2008, Mens 2010, Lukianova-Hleb 2013]. Hemozoin is a micro-crystalline heme compound produced by all *Plasmodium* spp. during the intraerythrocytic stage as they detoxify free heme derived from hemoglobin digestion [Francis 1997, Slater 1991, Fulton 1953]. The content of hemozoin increases as the parasite matures, thus constituting an optimal indicator of parasite maturation. Alongside infection diagnostics, the hemozoin-based detection of parasite maturation can be utilized in antimalarial drug development as well. The feasibility and efficiency of the concept has been recently demonstrated in a novel reagent-free drug sensitivity assay based on the flow cytometric detection of hemozoin [Rebelo 2013, Rebeleo 2015b].

As previously described, the rotating-crystal magneto-optical (MO) technique utilizes the linear dichroism and magnetic anisotropy highly specific to the hemozoin crystals to determine their concentration in liquid suspensions [Butykai 2013]. In our recent study the sensitivity of the method has been evaluated using ring and schizont stages from *P. falciparum in vitro* cultures where detection thresholds of 0.0008% and 0.0002% parasitemia have been found, respectively [Orban 2014]. These sensitivity results supplemented by other advantages of the MO method – label and reagent-free, cost-effective and quick – implied, that it's a suitable candidate for the detection of the infection as a laboratory tool in malaria research or, eventually, as an in-field diagnostic tool.

However, these previous experiments left the following questions about the possible limitations of the technique unanswered: i) possible false negative results in cases of infections with only very early ring stage parasites in the circulation containing little amounts of hemozoin and ii) the possibility of false positive detections due to the presence of hemozoin in the blood for extended periods of time after an infection has been cleared *in vivo*.

Here the performance of the method in monitoring the onset and progression of the blood stage infection of a malaria mouse model is presented. The aim of the study was to systematically investigate the abovementioned questions *in vivo* before engaging in complex and expensive field trials. Therefore, to mimic the natural infection under controlled conditions, mice were inoculated with sporozoites and the onset of the blood stage was assessed by different diagnostic tools to compare the MO method's performance to well-known techniques. The control methods used in the study were light microscopy, real-time PCR analysis and flow cytometry.

MO measurements during successful treatment of mice with *P. berghei* infections were also performed in order to gain insight into the clearance rate of hemozoin as the viable parasites disappear from the circulation; and to establish a time-frame after which the MO signal in successfully treated mice vanishes.

# Methods

**Animals, parasites and treatment**

In two independent experiment series (A and B), four and three BALB/c mice (Charles River, Spain) were infected with the transgenic *P. berghei* ANKA (259cl2) that constitutively expresses GFP during the whole life cycle. Sporozoites were obtained by disruption of the salivary glands of freshly dissected, infected female *Anopheles stephensi* mosquitoes and collected in DMEM (Dulbecco's Modified Eagle Medium from GIBCO). Mosquitoes were bred at the insect facility of the Instituto de Medicina Molecular. Each mouse was inoculated with 50 000 sporozoites.

In the infection experiments blood was collected every 5 hours from each mouse starting at the 48 hours post-infection (pi) time point in the case of the A series; and starting at 70 hours pi for series B, until the 80$^{th}$ hour in both cases. Afterwards, blood was collected daily until day 5 for both series.

In the treatment measurements (T series) five mice with severe *P. berghei* ANKA infections were treated by daily administration of 7 mg/ml of chloroquine for seven days. As the mice exhibited anemia on the first day of treatment as a consequence of the advanced infection, no further blood collection was performed until the administration of the third dose of chloroquine (day 2). On this day blood samples from 4 mice were analyzed again by the MO method and by light microscopy. Further blood sampling and analysis was performed on day 4, 5, 7, 10, 12, and 13, as shown in Figure 3. Microscopic observation of thin blood smears indicated that the treatment was successful and parasites have already been cleared from the blood stream on day 4, after the administration of five doses of chloroquine.

**Ethics statement**

This study was approved by the Ethical Committee of the Faculty of Medicine, University of Lisbon. All experiments involving animals were performed in compliance with the relevant laws and institutional guidelines. Animals were monitored daily and every effort was made to minimize suffering. Upon completion of experiments, mice were euthanized via administration of $CO_2$ followed by cervical dislocation.

**Magneto-optical measurements**

All reagents were purchased from Sigma Aldrich (St Louis, Mo, USA) unless stated otherwise. For the magneto-optical measurements 30 ul of blood was transferred from each mouse directly into 570 ul of lysis solution (0.066 V/V% Triton X-100 in 3 mM NaOH). The lysed sample was measured after 5 minutes, whilst the hemozoin crystals were liberated from the RBC's and from the parasites constituting a homogenous crystal suspension. MO measurements were performed using 450 ul from each lysed sample.

The prototype of the MO setup, as well as the underlying physical principles of the detection method, are described in a former study [Butykai 2013]. Briefly, the lysed sample, filled into a cylindrical sample holder, is inserted into the center of an assembly of permanent magnets arranged in a ring, which creates a strong uniform magnetic field (B = 1T) at the sample position and results in the co-alignment of the hemozoin crystals. When the magnetic ring is rotated, the co-aligned hemozoin crystals follow this rotation. During the measurement polarized light from a laser diode is transmitted through the sample in the direction perpendicular to the plane of the rotating magnetic field. The rotation of the co-aligned dichroic crystals gives rise to a periodic change in the transmitted intensity ($\Delta T$), which – divided by the time-averaged intensity (T) – corresponds to the measured MO signal ($\Delta T/T$ in %).

During the measurements described in this study, the rotational speed of the magnet was gradually increased from 1Hz to 50Hz, and the MO signal was recorded at each step. However, it has already been demonstrated that the best signal-to-noise ratio is observed in the range of 10-30 Hertz [Butykai 2013], and thus the MO signal value at 20 Hertz was chosen as a measure of the hemozoin content within the samples and is presented in the graphs showing the results of MO experiments.

The overall time required for the full MO measurement process is in the order of a few minutes. Consecutive measurements performed on the same lysed samples after 1, 10 and 24 hours of storage at 4°C show no changes of the MO signal within the margin of error, i.e., the standard deviation of three measurements performed consecutively on a given lysed sample.

The level of the uninfected baseline signal was determined prior to measuring the A and B experiment series by recording the MO signal of triplicate blood samples of 3 uninfected mice using the same protocol and instrumental settings as during the infection experiments. Additionally, one blood sample obtained from a control mouse at every post-infection (pi) time point was measured to check if any time-dependent variation of the baseline occurs. The MO signals of the real time control always fell within one standard deviation (SD) range of the previously established baseline. Thus, for clarity, we only plot the averaged baseline and its standard deviation in graphs summarizing the results of the MO measurements during infection and treatment (Figure 1B and Figure 3, respectively).

**Microscopic analysis**

Blood parasitemia was monitored by microscopic analysis of Giemsa-stained thin blood smears at each measured time point. Smears were fixed in absolute methanol and stained with 10% Giemsa-solution prepared in PBS1X.

The presence of parasites in the early time points, i.e., before 90 h post-infection was determined by light microscopic examination of 10-20 fields (~ 4500-9000 scanned RBCs) with 1000x magnification using a bright-field microscope (Leica, Solms, Germany) performed by two microscopists, independently. The number of scanned RBCs indicates that in samples identified as negatives, the highest possible parasitemia could be ~0.01-0.02%. At later time points percentages of infected red blood cells and approximate age distribution of the parasites was assessed by light microscopic examination of 5-10 fields (~ 2250-4500 RBCs), thus the lowest parasitemia values accurately quantifiable were in the order of 0.02-0.04%.

**DNA extraction and PCR analysis**

At the selected time points 5 µl of blood was collected from the tail vein into 200 µl of PBS 1X. DNA extraction was performed using the DNeasy Blood & Tissue Kit (Quiagen, USA), according to the manufacturer's instructions. Real-time PCR analysis was performed in duplicates using 2 µl of DNA and the iTaq Universal SYBR Green Supermix from Bio-Rad according to the manufacturer's instructions. Expression levels of 18 s rRNA were normalized against the housekeeping gene seryl-tRNA synthetase (PbANKA_061540). Gene expression values were calculated based on the ΔΔ Ct method. Primer pairs used were: PbA 18S rRNA: 5'

GGAGATTGGTTTTGACGTT TATGTG3' and 5'GGAGATTGGTTTTGACGTTTATGTG3'; PBANKA_061540: 5' ATTGCTCAACCTTATCAAACTG3' and 5'AGCCACATCTGAACAACCG3'.

**Flow cytometric measurements**

A volume of 5 µl of blood was collected and diluted in 1 ml of PBS 1X at each time point for each mouse. This blood suspension was analyzed in the CyFlow® Blue instrument (Partec, Munster, Germany), which is equipped with a 488 nm excitation laser, and has detectors for Forward Scatter (FSC), Side Scatter (SSC), green fluorescence – FL1 (BP 535/35 nm), orange fluorescence – FL2 (BP 590/50 nm) and red fluorescence – FL3 (LP 630 nm). For this study the setup was modified as described previously [Rebelo 2012, Rebelo 2015a]. Flow cytometry data were analyzed using FlowJo software (version 9.0.2, Tree Star Inc., Oregon, USA). Depolarizing events were defined in plots of side-scatter (SSC) versus depolarized-SSC as those with a signal above the background observed in the uninfected control. GFP positive cells were determined in green fluorescence (FL1) versus red fluorescence (FL3) plots. An uninfected control was used to define GFP positive cells, which consisted of the ones with green fluorescence levels above the uninfected control. The uninfected baseline of the whole measurement series was determined as the average and one standard deviation of the GFP/DSS positive events measured on the uninfected control mouse in all of the investigated time points.

**Results and discussion**

**Detection of the onset of blood stage infection using MO method, light microscopy, PCR and flow cytometry**

In order to determine the MO method's potential to detect the onset of the blood stage infection and to assess disease progression, the presence of intraerythrocytic parasites was monitored in seven infected and one uninfected control mouse in two independent experiment series (series A and B) in the time points indicated in Figures 1 and 2. Microscopic observation of Giemsa-stained thin blood smears, real-time PCR analysis and flow cytometry were used as comparison methods. The results of the first detection of circulating parasites by microscopy, the MO-method and q-PCR are shown in Figure 1A, 1B and 1C, respectively.

By the examination of Giemsa-stained thin blood films the first parasites were found in the smears of two mice of the B series at 70 hours post-infection; in all mice of the A series at the 75 h pi time-point; and only at 85 h pi in the smears of mouse B-3 (Figure 1A).

The MO signals (Figure 1B) measured on samples from the A series fall in the range of the uninfected baseline at the first three time points (48 h, 56 h and 61 h), but they clearly exceed the detection limit between the two measured time points of 61 and 66 hours, indicating hemozoin and thus parasite occurrence in the bloodstream. Similarly, the MO values already significantly differ from the level of the uninfected baseline at the first measured time point of 70 h pi in all mice of the B series. As a conclusion, the results presented in Figure 1B show that the MO measurements of the two series exhibit very similar tendencies and suggest, that circulating parasites can be detected by the MO method as early as about 63-64 hours pi, when mice are infected with 50 000 *P. berghei* ANKA sporozoites.

Real-time PCR analysis was performed on blood samples drawn at the first four post-infection time points from mice in the A series, and on the 70 h time point samples of mice in series B. Results in Figure 1C show that the difference in the magnitude of the PCR signals between the non-infected and infected samples becomes significant already at the 56 h pi time point and increases further at 61 and 70 h post-infection. These data are in good agreement with previous findings described by [Zuzarte-Luis 2014b] under similar experimental settings. According to the presented results the PCR could detect the first parasites in the circulation 56 hours after inoculation, confirming the merozoite egress and the onset of the blood stage of the infection with the expected high sensitivity.

Flow cytometry was also used to investigate the amount of parasites in the bloodstream of mice in series A. During the flow cytometric measurements two different parameters were assessed simultaneously (Figure 2A and 2B): i) fluorescence of GFP expressing parasites [Franke-Fayard 2004, Franke-Fayard 2008] and ii) depolarized side scattering (DSS) properties of the hemozoin containing red blood cells [Rebelo 2012, Rebelo 2015a]. An inherent advantage of flow cytometry is that parasitemia counts are the directly determined quantities in both settings, unlike in the case of MO and PCR that yield indirect measures of parasite burden. However, these two values may implicitly differ as fluorescence detection measures all DNA-containing parasites above a certain sensitivity-level, while by the analysis of depolarized side scattering, only the fraction of parasitized RBCs containing sufficient amounts of hemozoin crystals can be detected [Rebelo 2012].

The results of the DSS measurements (Figure 2B) of the infected mice are scattered within the range of the uninfected baseline (see Methods) until the time point of 90 h post-infection, when the percentage of the depolarizing cells starts to increase monotonously in all infected mice. The first positive confirmation of the blood stage infection by DSS measurements, i.e. when the signal magnitude starts to differ significantly from the uninfected baseline, is 90 hours post-infection in all three infected mice. At this point the percentages of hemozoin containing parasitized RBCs are 0.062%, 0.045% and 0.039% for mice A-1, A-2 and A-4, respectively.

The parasite burdens detected via fluorescence (Figure 2A) were indeed higher at any time point than the values measured by DSS, indicating thus a better sensitivity and the possibility of an earlier positive diagnosis. However, as the baseline signal of the GFP measurements was also proportionally higher than the baseline of the DSS measurements, the final sensitivity was only moderately better exposing the infection at 85 h pi in all three studied mice. The corresponding parasitemia values at this time point were: ~0.11%, ~0.14% and ~0.07% for mouse A-1, A-2 and A-4, respectively.

**Monitoring the progression of the blood stage infection by the MO method**

After establishing the time points of the first positive parasite detection by the exposed methods, the progression of the infection at later time points was monitored by light microscopy, flow cytometry and MO measurements.

Assessment of parasitemia by microscopy in time points after 90 hours pi suggests (Figure 2C), that all infected mice developed standard blood infections with parasitemias ranging between 0.12% and 0.7% at day 4 post-infection (90-100 h pi time points). These results are in good agreement with Ref. [Ploemen 2009], where experiments were performed with similar initial sporozoite loads.

As the MO measurements confirmed the onset of the blood stage already at 66 h pi, the course of the infection could be monitored quantitatively from this early stage, contrary to light microscopy and flow cytometry, where reliable parasite detection was possible only 10 and 20 hours later, respectively (Figure 2B).

The MO signals of all three infected mice from series A show the same profile after they surpass the level of the uninfected baseline: the MO signal increases in a rapid and monotonous manner, interrupted by a distinct drop at 76h. Furthermore, a drop in the increase rate can be identified between 90h and 100h post-infection. The same behavior is reproduced by mice in series B, with only slight variations in the times of the drops in the signal, observable at 80h, 95h and 125h. While the drops at the early time points (76 h and 80 h) are significant, showing a decrease in the MO signal of 65% and 57% for series A and B, respectively, at later time points they are much less pronounced.

In general it can be concluded that the overall time dependence of the MO signal is very similar in all studied mice except for the 4-5 hours delay of the members of series B compared to series A. These results suggest that the early blood stage of the infection followed quite similar courses in all studied mice – as supported by the results of flow cytometry and microscopy – and that this progression could be reliably monitored by the MO measurements. For the interpretation of the time dependence of the MO signal it has to be noted that the MO method measures the overall hemozoin content in the peripheral circulation, which includes crystals present within the parasites at the moment of blood sampling, free in circulation and/or inside phagocytic cells. Consequently, changes in the signal magnitude in the early phases of blood stage development are likely to result from two dynamic processes: i) the continuous production of hemozoin by the circulating parasites, which increases the MO signal, and ii) the clearance of free hemozoin or hemozoin-containing phagocytes from circulation leading to a decrease in the MO signal. If these two processes have comparable production (clearance) rates in a synchronous infection at a given parasite density, the MO signal is expected to pursue the following tendency: (i) monotonous increase from the beginning of the first cycle (ii) reaching a maximum at the end of the cycle when mature schizonts have maximal hemozoin production rate (iii) decrease due to the rupture of iRBCs and the subsequent hz clearance (iv) turnover and increase due to the hemozoin production of the new generation of parasites.

Correspondingly, the significant drop in the MO signal observed at around 76-80 hours for all mice suggests that a synchronous schizont rupture occured in the previous hours. According to literary data the length of the life cycle in the case of *P. berghei* parasites is 22-23 hours [Janse 1995] and merozoite egress from hepatocytes is expected at around 43-52 hours after sporozoite injection [Janse 1995, Liehl 2013]. Therefore, the first cycle in the blood is expected to end between 65-75 hours pi. This timeframe coincides well with the peak of the MO signal observed at ~70-75 h and followed by the drop at 76-80 hours in both series. This derived time course of the first cycle was also confirmed by the microscopic observation of Giemsa-stained smears where the first parasites found at 70 and 75 hours were indeed late trophozoites and schizonts, respectively. These observations indicate that the first parasites detected by the MO method around 63-64 hours pi were ~10-14 h old rings of the first, synchronous life cycle. The drops in the increase rate observed in the MO signals at later time points (~109 h in mice A and ~ 100 h in mice B) also correspond to the 22-23 h periodicity of the life cycle suggesting that a partial synchronicity is still preserved at the end of the second cycle. However, the decreasing relative magnitude of the drops confirms that this synchronicity is less and less preserved, and eventually lost, as the infection progresses, which is indeed a well-known characteristic of *P. Berghei* infections [Janse 1995].

**Assessment of the parasitemia level in the first intraerythrocytic cycle by the MO method**

In our previous studies we demonstrated that the MO signal is directly proportional to the concentration of both synthetic β-hematin [Butykai 2013] and natural hemozoin crystals produced by *in vitro P. falciparum* parasite cultures [Orban 2014]. Using a conversion factor between the concentration of synthetic hemozoin crystals and the magnitude of the MO signal [Orban 2014] the amount of circulating hemozoin can be determined at any time point of the infection. Moreover, based on data available in the literature regarding conversion rates of hemoglobin into hemozoin in *Plasmodium* species [Francis 1997, Weissbuch 2008, Gligorijevic 2006], the parasitemia level and age distribution that is sufficient to produce a given amount of hemozoin can be calculated. Using these two conversion factors, we give a rough estimate for the initial parasitemia value at the end of the first synchronous erythrocytic cycle. In the case of series A the average of the MO values at the 71 hour time point, i.e., at the end of the 1$^{st}$ cycle, when most of the circulating parasites are estimated to be schizonts is $\Delta T/T = 5.6 \pm 2*10^{-3}$ %. According to our previous report [Butykai 2013], this would correspond to a hemozoin concentration of 0.08 ± 0.029 ng/μl in the measured crystal suspension that was produced by a 20-fold dilution of the collected whole blood sample. Assuming that by the end of their life cycle the schizonts have converted roughly 80% of the host cell's hemoglobin into hemozoin, and using standard hemoglobin and RBC concentration values for BALB/c mice [Russell 1951], the parasitemia level at the end of the first erythrocytic cycle is estimated to be 0.0018 ± 0.0007%. This estimate for the parasitemia level at the end of the first erythrocytic cycle is in good agreement with the first direct light-microscopy count of ~0.15% at 90 hours pi (beginning of the third erythrocytic cycle) using an approximately 8-fold multiplication rate upon invasion in accordance with data available in the literature [Janse 2003, Janse 2006].

To support the consistency of the above estimations and the robustness of the hemozoin concentration measurements via the MO method, a reversed calculation can be applied for the MO values measured at 66 hour pi when circulating parasites are assumed to be 10-14 hours old rings, at the previously established ~0.0018% parasitemia level. The averaged MO value of $\Delta T/T = 1.9 \pm 0.7 \times 10^{-3}$ % corresponds to a hemozoin concentration of 0.027 ± 0.01 ng/µl. In order to produce this amount of hemozoin parasites have had to convert 25-30% of the host cell's hemoglobin, which is in good agreement with data found in the literature [Francis 1997, Weissbuch 2008, Gligorijevic 2006]. This correspondence confirms that the MO signal can be consistently related to the parasitemia level and the course of the infection can be traced via the precise measurement of *in vivo* hemozoin production.

**Monitoring parasite clearance by the MO method**

The ability of the MO method to follow parasite clearance during and after treatment, and the persistence of positive test results were also investigated.

In this study 5 mice with severe *P. berghei* infection were treated by daily administration of chloroquine for seven days as described in the 'Methods' section. The MO signal was measured both during the treatment and for the nine consecutive days after treatment to study i) the correlation between parasite clearance and the magnitude of MO signal and ii) the timescale over which the MO signal is reduced to the level of the uninfected baseline following a successful treatment.

The first blood samples were collected from mice T-1, T-3 and T-4 on the first day of treatment (day 0 in Figure 3), and according to the microscopic observation of Giemsa-stained thin blood smears, their parasitemia values were 14.7%, 20% and 10.8%, with corresponding MO values of 3.76, 5.2 and 2.78, respectively (Figure 3). As the mice exhibited severe anemia on the first day of treatment, no further blood collection was performed until the administration of the third dose of chloroquine (day 2). Further blood sampling and analysis was performed on days indicated in Figure 3.

Similarly to the infection experiments, the MO signals during the treatment of all studied mice exhibit good correlation and follow very similar, monotonously decreasing trends (Figure 3). The significant decrease in parasitemia caused by the administration of the first three doses of chloroquine is reflected in the rapid (approximately two orders of magnitude) decrease of the MO signal observed between day 0 ($\Delta T/T=3.9 \pm 1,2$ %) and day 3 ($\Delta T/T=0.014 \pm 0.004$%). The signals monotonously decrease during the consecutive days in as a result of parasite clearance, however, the MO values still don't reach the level of the uninfected baseline on day 4, when there were no viable parasites in the circulation observable by light microscopy anymore. This suggests that the MO signals measured from day 4 to day 7 are likely to originate from hemozoin crystals released from ruptured schizonts being still present in the circulation either freely or inside phagocytes [Frita 2011, Boura 2013].

The presence and the kinetics of Hz-containing phagocytes in peripheral blood of malaria patients has been addressed in several studies [Lyke 2003, Hanscheid 2008, Kremsner 2009]. The decrease observed by the MO method from day 4 to day 7 may reflect the clearance rate of hemozoin from circulation in mice: in the course of 72 hours the hemozoin concentration dropped from $3.6 \pm 0.73$ ng/ml to $0.59 \pm 0.25$ ng/ml. These data fit with the findings of the referred clinical study, where a median half-life of 216 hours for Hz-containing monocytes was reported for *P. falciparum* infections [Day 1996] and also with the observations of mid- and long-term hemozoin kinetics reported for in vivo *P. Berghei* infections [Frita 2011]. In the latter study 7 days after parasite clearance ~1.7% of the phagocytes were found to contain hemozoin granules.

By day 10 the MO signals in all treated mice decreased to the level of the uninfected baseline. Indeed, the signal remained at this level for the consecutive two days, confirming the absence of the infection and yielding true negative diagnostic results.

## Conclusions

The rotating-crystal magneto-optical detection method has demonstrated excellent sensitivity to detect low concentrations of synthetic hemozoin crystals [Butykai 2013] and low parasite densities in *P. falciparum in vitro* cultures [Orban 2014], suggesting its potential to be developed into a diagnostic tool. Here, using a rodent model of malaria the *in vivo* sensitivity of the method was investigated.   In the presented study the MO method was able to detect the first generation of blood stage parasites at the ring stage between 61 and 66h hours after sporozoite injection. These values demonstrate a better performance of the MO method than the ~70-75 hour pi observation of the first parasites by standardly performed light microscopy, and is exceeded only by the laboratory-grade q-PCR method, which was able to detect parasites 56 h pi.

Even though *P. berghei* infections are regarded as suitable systems for modeling the blood stage of *P. vivax* human infections [Zuzarte-Luis 2015], the direct extrapolation of results obtained from animal models to human infections is not straightforward, especially in the case of *P. falciparum*.  In these infections, erythrocytic parasite stages later than early trophozoites tend to cytoadhere to the endothelium, thus the circulating viable parasites are the young forms containing little or no hemozoin. Clearly, this scenario cannot be addressed using a *P. berghei* infection model, however, the current study yielded two results also with important implications to of *P. falciparum* infections: (i) the first generation of parasites at the mid-ring stage were already detectable; (ii) the MO results both during infection and treatment show, that there is a substantial amount of hemozoin crystals circulating in the blood stream for a few days after schizont rupture. These observations indicate that the MO method has real potential to identify human *P. falciparum* infections with good sensitivity, either by the hemozoin being released from the sequestered schizonts or directly via the lesser amounts of hemozoin produced by freely circulating mid-stage rings.

We performed treatment experiments on infected mice in order to address the issue of possible false positive outcomes after successful treatment of infected individuals. The results of these measurements revealed, that the MO signal returns to the level of the uninfected baseline after a few days of parasite clearance in mice, suggesting a similar scenario in the case of human infections.

While studies on field isolates are required for the final evaluation of the method for human infections, we believe that the present study provides a solid basis for the implementation of such in-field tests.

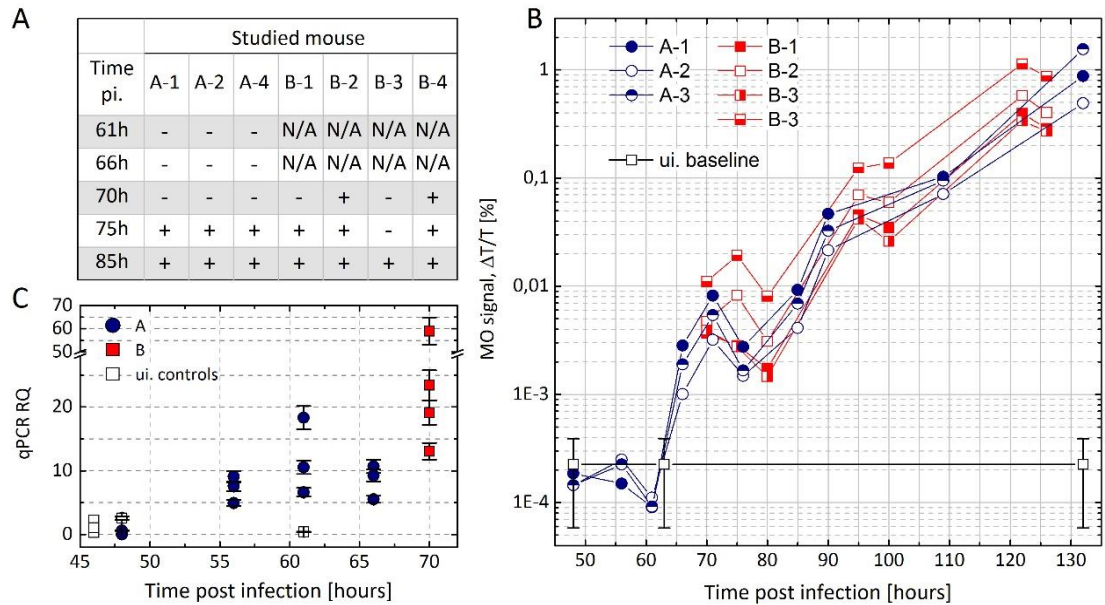

**Figure 1 Determination of the onset of blood stage infection using light microscopy, magneto-optical method and q-PCR. A:** Results of the microscopic examination of Giemsa-stained thin blood smears for each mouse at the presented time points. + signals indicate that at least one parasite was found in the given smear; - signals denote the complementary case. **B**: The results of the MO measurements. Each circle/square represents the MO signal for a given mouse in series A/B at a given time point after sporozoite injection. The lines are guides for the eye. The open black squares represent the average uninfected baseline signal determined prior to the A and B measurement series (for details see Methods section). The error bars correspond to the standard deviation of these baseline measurements. The MO signal of all mice in series A exceeds the value of $\Delta T/T = 3.9 \times 10^{-4}\%$ (ui. baseline + 1* standard deviation) between the measured time points of 61h and 66h. **C**: The results of the q-PCR measurements. The blue circles/red squares represent the PCR values of mice belonging to the A/B series. The error bars represent the standard deviation of technical duplicates. The open squares represent measurements on three uninfected controls and the real-time control mouse at 61 h.

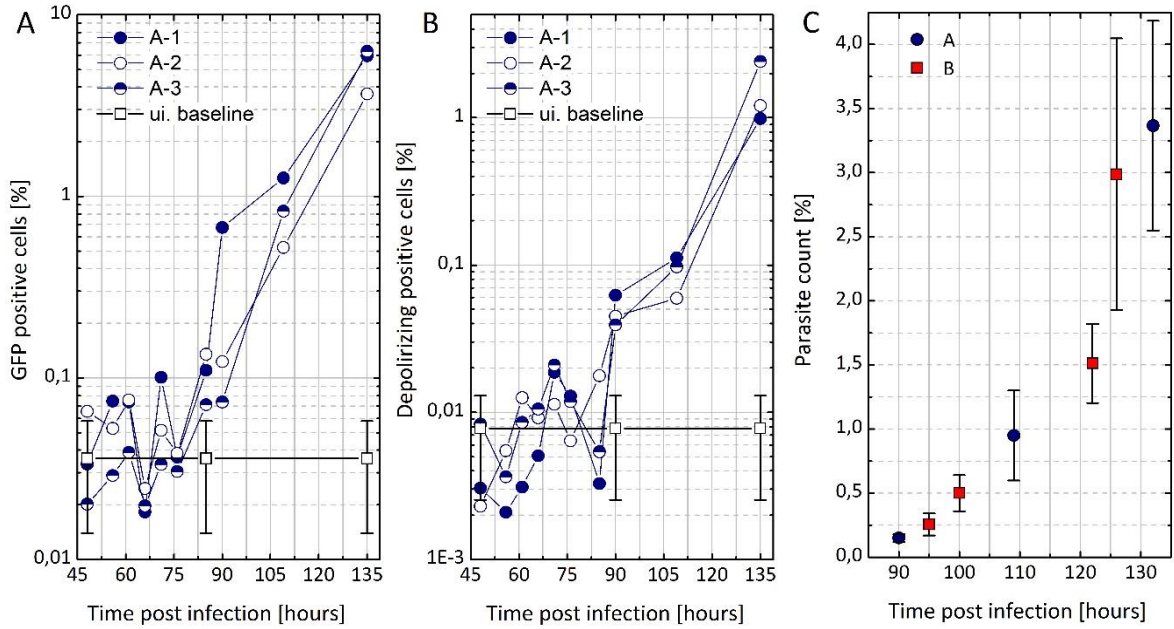

**Figure 2 Monitoring parasitemia during the progression of the infection via flow cytometry and light microscopy. A:** The results of the GFP measurements. The circles represent percentage of GFP positive events in the total population of RBC counts for a given mouse in series A. The lines are guides to the eye. The open squares represent the averaged GFP value measured on 9 uninfected control samples of the real-time control mouse. The GFP signal of all mice in series A exceeds the value of 0.058% (ui. baseline + 1* standard deviation) at the 85 h time point **B:** The results of the DSS measurements. The circles represent percentage of DSS positive events in the total population of RBC counts for a given mouse in series A. The lines are guides to the eye. The open squares represent the averaged DSS value measured on 9 uninfected control samples of the real-time control mouse. The DSS signal of all mice in series A exceeds the value of 0.013% (ui. baseline + 1* standard deviation) at the measured time point of 90 h. **C**: Averaged parasitemia values determined by light microscopy after 90 h pi at the inspected time points of series A (blue circles) and B (red squares).

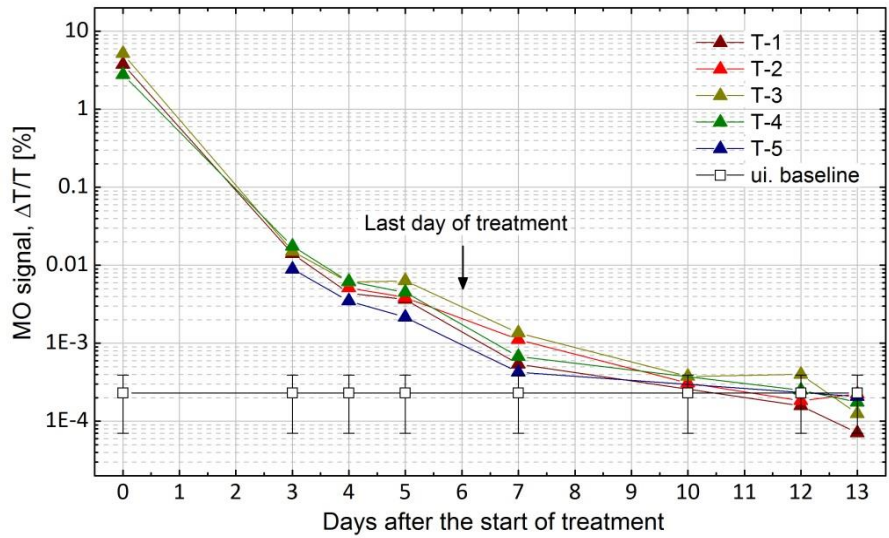

**Figure 3 Monitoring parasite clearance during treatment by the MO method.** The MO values of five mice were measured during and post treatment (T-series). Each colored triangle represents the MO value of a given mouse on a given day after the start of the treatment. The lines are guides to the eye. The open black squares and bars represent the averaged MO signal of 3 uninfected control samples measured in technical triplicates and their standard deviation, respectively. The averaged MO value of the treated mice reaches the level of the uninfected baseline on day 10 and stays within this level in the two consecutive measurements.

# References

[1] Zuzarte-Luis V, Mota MM, Vigário AM: **Malaria infections: What and how can mice teach us.** J Immunol Methods 2014, **410**:113-122.

[2] Mota MM, Thathy V, Nussenzweig RS, Nussenzweig V: **Gene targeting in the rodent malaria parasite Plasmodium yoelii**. Mol Biochem Parasitol 2001, **113**:271-278. doi:10.1016/S0166-6851(01)00228-6.

[3] Khan ZM, Vanderberg JP: **Role of host cellular response in differential susceptibility of nonimmunized BALB/c mice to Plasmodium berghei and Plasmodium yoelii sporozoites**. Infect Immun 1991, **59**:2529-2534.

[4] White NJ, Turner GD, Medana IM, Dondorp AM, Day NP: **The murine cerebral malaria phenomenon**. Trends Parasitol 2010, **26**:11-15.

[5] Craig AG, Grau GE, Janse C, Kazura JW, Milner D, Barnwell JW, Turner G, Langhorne J.: **The role of animal models for research on severe malaria**. PLoS Pathog 2012, **8**:e1002401.

[6] Frean J: **Improving quantitation of malaria parasite burden with digital image analysis**. Trans R Soc Trop Med Hyg 2008, **102**:1062-1063.

[7] Valkiunas G, Iezhova TA, Krizanauskiene A, Palinauskas V, Sehgal RN, Bensch S: **A comparative analysis of microscopy and PCR-based detection methods for blood parasites**. J Parasitol 2008, **94**:1395–1401.

[8] Taylor SM, Juliano JJ, Trottman PA, Griffin JB, Landis SH, Kitsa P, Tshefu AK, Meshnick SR: **High-throughput pooling and real-time PCR-based strategy for malaria detection**. J Clin Microbiol 2010, **48**:512–519.

[9] Lelliott PM, Lampkin S, McMorran BJ, Foote SJ, Burgio G: **A flow cytometric assay to quantify invasion of red blood cells by rodent Plasmodium parasites in vivo**. Malar J 2014, **13**:100

[10] Jimenez-Diaz MB, Mulet T, Gomez V, Viera S, Alvarez A, Garuti H, Vazquez Y, Fernandez A, Ibanez J, Jimenez M, Gargallo-Viola D, Angulo-Barturen I: **Quantitative measurement of Plasmodium-infected erythrocytes in murine models of malaria by flow cytometry using bidimensional assessment of SYTO-16 fluorescence**. Cytometry A 2009, **75**:225–235.

[11] Malleret B, Claser C, Ong AS, Suwanarusk R, Sriprawat K, Howland SW, Russell B, Nosten F, Renia L: **A rapid and robust tri-color flow cytometry assay for monitoring malaria parasite development**. Sci Rep 2011, **1**:118.

[12] Franke-Fayard B, Trueman H, Ramesar J, Mendoza J, van der Keur M, van der Linden R, Sinden RE, Waters AP, Janse CJ: **A Plasmodium berghei reference line that constitutively expresses GFP at a high level throughout the complete life cycle**. Mol Biochem Parasitol 2004, **137**:23–33.

[13] Franke-Fayard B, Djokovic D, Dooren MW, Ramesar J, Waters AP, Falade MO, Kranendonk M, Martinelli A, Cravo P, Janse CJ: **Simple and sensitive antimalarial drug screening in vitro and in vivo using transgenic luciferase expressing Plasmodium berghei parasites.** Int J Parasitol 2008, **38**:1651-1662.

[14] Zuzarte-Luis V, Sales-Dias S, Mota MM: **Simple, sensitive and quantitative bioluminescence assay for determination of malaria pre-patent period**. Malar J 2014, **13**:15.

[15] Ono T, Tadakuma T, Rodriguez : **Plasmodium yoelii yoelii 17XNL constitutively expressing GFP throughout the life cycle.** Exp Parasitol 2007, **115**:310-313.

[16] and that they must be kept under constant drug selection to preserve a pure luciferase expressing line

[17] Rebelo M, Shapiro HM, Amaral T, Melo-Cristino J, Hänscheid T: **Haemozoin detection in infected erythrocytes for Plasmodium falciparum malaria diagnosis—Prospects and limitations** Acta Trop. 2012, **123**:58– 61

[18] Rebelo M, Tempera C, Bispo C, Andrade C, Gardner R, Shapiro HM, Hänscheid T: **Light depolarization measurements in malaria: A new job for an old friend.** Cytometry A. 2015, **87:** 437–445

[19] Karl S, David M, Moore L, Grimberg BT, Michon P, Mueller I, Zborowski M Zimmerman PA: **Enhanced detection of gametocytes by magnetic deposition microscopy predicts higher potential for plasmodium falciparum transmission.** Malar J 2008, **7**:66.

[20] Mens PF, Matelon RJ, Nour BYM, Newman DM, Schallig H: **Laboratory evaluation on the sensitivity and specificity of a novel and rapid detection method for malaria diagnosis based on magneto-optical technology (MOT)**. Malar J 2010, **9**:207.

[21] Lukianova-Hleb EY, Campbell KM, Constantinou PE, Braam J, Olson JS, Ware RW, Sullivan DJ Jr, Lapotko DO: **Hemozoin-generated vapor nanobubbles for transdermal reagent- and needle-free detection of malaria.** Proc Nat Acad Sci USA 2013, **11**:900-905.

[22] Francis SE, Sullivan DJ, Goldberg DE: **Hemoglobin metabolism in the malaria parasite plasmodium falciparum**. Annu Rev Microbiol 1997, **51**:97–123.

[23] Hänscheid T, Egan TJ, Grobusch MP: **Haemozoin: from melatonin pigment to drug target, diagnostic tool, and immune modulator**. Lancet Infect Dis 2007, **7**:675–685.

[24] Rebelo M, Sousa C, Shapiro HM, Mota MM, Grobusch MP, Hänscheid T**: A novel Flow Cytometric hemozoin detection assay for real-time sensitivity testing of Plasmodium falciparum**. PLoS ONE 2013 **8**:e61606.

[25] Rebelo M, Tempera C, Fernandes J, Grobusch MP, Hanscheid T: **Assessing anti-malarial effcts ex vivo using the haemozoin detection assay**. Malar J 2015, **14**:140.

[26] Butykai A, Orban A, Kocsis V, Szaller D, Bordacs S, Tátrai-Szekeres E, Kiss LF, Bóta A, Vértessy BG, Zelles T, Kézsmárki I: **Malaria pigment crystals as magnetic micro-rotors: key for high-sensitivity diagnosis**. Sci Rep 2013, **3**:1431.

[27] Orban A, Butykai A, Molnár A, Pröhle ZS, Fülöp G, Zelles T, Forsyth W, Hill D, Müller I, Schofield L, Rebelo M, Hänscheid T, Karl S, Kézsmárki I: **Evaluation of a novel magneto-optical method for the detection of malaria parasites.** PLoS ONE 2014, **9**:e96981.

[28] Ploemen IHJ, Prudêncio M, Douradinha BG, Ramesar J, Fonager J, Gemert GJ, Luty AJF, Hermsen CC, Sauerwein RW, Baptista FG, Mota MM, Waters AP, Que I, Lowik CWGM, Khan SM, Janse CJ, Franke-Fayard B: **Visualisation and quantitative analysis of the rodent malaria liver stage by real time imaging.** PLoS ONE 2009, **4**:11.

[29] Janse CJ, Waters AP:Plasmodium berghei: **The application of cultivation and purification techniques to molecular studies of malaria parasites.** Parasitol Today 1995, **4**:138–143.

[30] Liehl P, Zuzarte-Luis V, Chan J, Zillinger T, Baptista F, Carapau D, Konert M, Hanson KK, Carret C, Lassnig C, Muller M, Kalinke U, Saeed M, Chora AF, Golenbock DT, Strobl B, Prudencio M, Coelho LP, Kappe SH, Superti-Furga G, Pichlmair A, Vigario AM, Rice CM, Fitzgerald KA, Barchet W, Mota MM: **Host-cell sensors for plasmodium activate innate immunity against liver-stage infection**. Nat Med 2013, **20**: 47–53.

[31] Francis SE, Sullivan DJ, Goldberg DE: **Hemoglobin metabolism in the malaria parasite plasmodium falciparum.** Annu Rev Microbiol 1997, **51**:97–123.

[32] Weissbuch I, Leiserowitz L: **Interplay between malaria, crystalline hemozoin formation, and antimalarial drug action and design.** Chem Rev 2008, **108**:4899–4914.

[33] Gligorijevic B, McAllister R, Urbach JS, Roepe PD: **Spinning disk confocal microscopy of live, intraerythrocytic malarial parasites. 1. Quantification of hemozoin development for drug sensitive versus resistant malaria**. Biochemistry 2006, **45**:12400-10.

[33.3] Janse CJ, Haghparast A, Speranca MA, Ramesar J, Kroeze H, Portillo HA, Waters AP: **Malaria parasites lacking eef1a have a normal S/M phase yet grow more slowly due to a longer G1 phase.** Mol Microbiol 2003, **50**:1539–51.

[33.5] Janse CJ, Ramesar J, Waters AP: **High-efficiency transfection and drug selection of genetically transformed blood stages of the rodent malaria parasite Plasmodium berghei.** Nat Protoc 2006, **1**: 346–356.

[34] Frita R, Rebelo M, Pamplona A, Vigario AM, Mota MM, Grobusch MP, Hänscheid T: **Simple flow cytometric detection of haemozoin containing leukocytes and erythrocytes for research on diagnosis, immunology and drug sensitivity testing.** Malar J 2011, **10**:74.

[35] Boura M, Frita R, Góis A, Carvalho T, Hänscheid T: **The hemozoin conundrum: is malaria pigment immune-activating,inhibiting, or simply a bystander?** Trends in Parasitology 2013, **10**:469–476.

[36] P. G. Kremsner, C. Valim, M. A. Missinou et al., "**Prognostic value of circulating pigmented cells in African children with malaria**," Journal of Infectious Diseases, vol. 199, no. 1, pp. 142–150, 2009.

[37] K. E. Lyke, D. A. Diallo, A. Dicko et al., "**Association of intraleukocytic Plasmodium falciparum malaria pigment with disease severity, clinical manifestations, and prognosis in severe malaria**," The American Journal of TropicalMedicine and Hygiene, vol. 69, no. 3, pp. 253–259, 2003.

[38] T. Hanscheid, M. L¨angin, B. Lell et al., "**Full blood count and haemozoin-containing leukocytes in children with malaria: diagnostic value and association with disease severity**," Malaria Journal, vol. 7, article 109, 2008.

[39] N. P. J. Day, T. D. Pham, T. L. Phan et al., "**Clearance kinetics of parasites and pigment-containing leukocytes in severe malaria**," Blood, vol. 88, no. 12, pp. 4694–4700, 1996.

[40] Russell, E.S., E.F. Neufeld, and C.T. Higgins. 1951. **Comparison of normal blood picture of young adults from 18 inbred strains of mice**. Proc. Soc. Exp. Biol. Med. **78:** 761-766.